\documentclass[preprint2]{aastex}

\newcommand{\sfru}{M$_{\odot}$yr$^{-1}$}
\newcommand{\msun}{M$_{\odot}$}

\slugcomment{Accepted for ApJ}

\shorttitle{Star formation in LSB galaxies}
\shortauthors{Boissier et al.}

\begin{document}

\title{GALEX observations of Low Surface Brightness Galaxies: \\
       UV color and star formation efficiency\altaffilmark{7}}

\author{S. Boissier\altaffilmark{1,2}, 
A. Gil de Paz\altaffilmark{3}, 
A. Boselli\altaffilmark{1,2}, 
V. Buat\altaffilmark{1}, 
B. Madore\altaffilmark{4}, \\
L. Chemin\altaffilmark{5},
C. Balkowski\altaffilmark{5}, 
P. Amram\altaffilmark{1}, 
C. Carignan\altaffilmark{6},
W. van Driel\altaffilmark{5}
}

\altaffiltext{1}{Laboratoire d'Astrophysique de Marseille, Traverse  du Siphon-Les trois Lucs, BP8-13376 Marseille Cedex 12, France}
\altaffiltext{2}{CNRS, UMR~6110}
\altaffiltext{3}{Dept. de Astrofisica y CC. de la Atmosfera, Universidad Complutense de Madrid, Avda. de la Complutense, s/n Madrid, E-28040 Spain}
\altaffiltext{4}{Observatories of the Carnegie Institution of Washington, 813 Santa Barbara Street Pasadena, California 91101, USA}
\altaffiltext{5}{Observatoire de Paris, GEPI, CNRS and Universit\'e Paris 7, 5 Place Jules Janssen, 92195 Meudon Cedex, France}
\altaffiltext{6}{Universit\'e de Montr\'eal, CP 6128, Succ centre-ville, Montr\'eal, P. QC Canada, H3C-3J7}
\altaffiltext{7}{Based on observations made with the NASA Galaxy Evolution Explorer GALEX is operated for NASA by the Califonia Institute of Technology under NASA contract NAS5-98034}

\begin{abstract}
  We present GALEX UV observations of a sample of Low Surface
  Brightness (LSB) galaxies for which HI data are available, allowing
  us to estimate their star formation efficiency.
  We find that the UV light extends to larger radii than the optical
  light (some galaxies, but not all, look similar to the recently
  discovered XUV-disk galaxies).
  Using a standard calibration to convert the UV light into star
  formation rates, we obtain lower star formation efficiencies in LSB
  galaxies than in high surface brightness galaxies by about one order
  of magnitude.
  We show however that standard calibrations may not apply to these
  galaxies, as the FUV-NUV color obtained from the two GALEX bands
  (FUV and NUV; $\lambda_{\mathrm{eff}}$= 1516 and
  2267 \AA, respectively) is redder than expected for star forming galaxies.  
   This color can be interpreted
  as a result of internal extinction, modified Initial Mass Function
  or by star formation histories characterized by bursts followed by
  quiescent phases. Our analysis favors this latter hypothesis.
\end{abstract}


\section{Introduction}

In the very last years, outskirts of galaxies and low density regions
have been the subject of a renewed interest, especially after the
discovery of extended ultraviolet (XUV) disks in nearby galaxies with GALEX
\citep{gildepaz05, thilker05}.
The works by \citet{zaritsky07} and \citet{thilkerxuv} suggests that
about 30 \% of disk galaxies do present some level of XUV emission.
\citet{boissier06} shown that the UV reveal stellar formation in the
outskirts of normal galaxy disks, including regions beyond
the usual ``threshold'' for star formation derived from
H$\alpha$ azimuthally-averaged profiles \citep{martin01}.
The ultraviolet GALEX observations also allowed to estimate the amount
of star formation in
low density regions in the case of interactions. For instance, 
\citet{boselli05} discovered in NGC4438 a tidal tail detected only
in UV, and estimated the age and the amplitude of the burst
induced by the interaction. Similarly \citet{boquien07} studied star 
formation in the intergalactic medium around NGC~5291, expelled from 
parent galaxies after a collision.

As noted by \citet{thilkerxuv}, a link between the XUV phenomenom,
star formation in low density regions and Low Surface Brightness
(LSB) galaxies is very likely. 
According to a hand waving definition, a disk galaxy should be
considered as a LSB galaxy if its disk central face-on surface brightness in
the B band is well below the typical Freeman value of 21.65 mag
arcsec$^{-2}$ \citep[see e.g.][]{bothun97}. LSB galaxies include quite
different populations ranging from dwarf galaxies (faint and relatively compact
objects) to disk galaxies, and even ``giant'' disk galaxies with
scale-lengths larger than 5 kpc and masses comparable to the more
massive spirals \citep{oneil98}. Various samples of ``LSB galaxies''
do not always trace the same population, and one should take this fact
into account.
In this paper, we will distinguish ``massive'' and ``low-mass'' LSB galaxies
according to the HI total mass since we have this quantity for every object
in our sample. 

%
LSB have in general faint surface brightnesses, 
blue colors \citep[see e.g.][]{deblok95}
large amounts of neutral gas \citep{oneil98}, 
low metallicity \citep{mcgaugh94}.
Similarly, XUV regions have low metallicities \citep{gildepaz07} and 
are found in galaxies that are systematically more gas-rich than
the general field galaxy population \citep{thilkerxuv}. 
Star formation in XUV disks and LSB galaxies is 
thus likely to share some characteristics.

Several models to explain the existence and properties of LSB galaxies
have been proposed
\citep[e.g.][]{jimenez98,vandenhoek00,gerritsen99,boissier03}. All of
them rely on the inclusion of a recipe for star
formation, often characterized by a lower efficiency with respect to
High Surface Brightness (HSB) galaxies, either related to structural
parameters (larger size and lower densities) or to metallicity.
Characterizing the star formation in LSB galaxies is thus an important
step that will bring constraints on their modeling.
Star formation rates in LSB galaxies were derived from their optical
properties \citep[e.g.][]{mcgaugh94b,vandenhoek00}, and a few attempts
to measure them are found in \citet{burkholder01,vanzee97,oneil07}.
None of these studies are based on UV data while GALEX has shown the
interest of the ultraviolet domain to reveal star formation in low
density regions.

On the basis of these considerations, we observed with GALEX a sample of
LSB galaxies for which HI data are available.

This paper presents the results of a first analysis of these
observations (described in section 2), including the study of the
spatial distribution of the UV light (section 3) and the determination of
star formation rates and efficiencies in LSBs (section 4). In section
5, we discuss the color of our LSB galaxies, followed (section 6) by
the consequences on the validity and interpretation of the star
formation rates that we derived and on the star formation history
of such galaxies.  This discussion may apply to other
cases than LSB galaxies, like for instance XUV-disk galaxies,
considering the similarities they share.

\section{Description of our sample and data}

\subsection{LSB sample}

In order to study the transformation of gas to stars in LSB galaxies,
we compiled a list of such galaxies with a measured HI mass
\citep[taken from][]{vanderhulst93,deblok96,matthews01,chung02}.  We
obtained GALEX Guest Investigator imaging (proposal 67, cycle1) in order to determine
the UV emission of 10 of these objects, being ``massive'' or
giant LSB galaxies \citep[e.g.][]{matthews01}. We present in this
paper the observations actually performed with GALEX to date for this
proposal.
We also include in our analysis other LSB galaxies with HI
measurements that were also observed by GALEX in the framework
of various surveys, and for which the UV data are publicly
available. Although the sample is not complete in any sense, 
it includes 18 galaxies ranging from ``Dwarf'' to ``Giant'' LSB 
galaxies ($-12.12>M_B>-22.90$ and $7.11<log(M(HI))<10.66$).
Table \ref{tabsample} gives the position and basic information for our galaxies,
taken from the NED database and the literature. 
The positions and position angles of a few galaxies were changed to
match our GALEX imaging data.

\clearpage
\begin{deluxetable}{l r r r r r r r r } 
\tablecaption{Basic properties of the sample\label{tabsample}} 
\startdata 
        Name     & RA      & DEC     & (2a)  & (2b)  & PA  & Distance & log(M(HI)) & $M_B$  \\ 
                 & (deg)   & (deg)   & (arcmin) & (arcmin) & (deg) & (Mpc)  & $M_{\odot}$ & (mag)  \\  
        UGC00568 &  13.787 &  -1.046 &   1.3 &   0.9 &   0 & 190.0 &  10.25 & -21.53 \\ 
        UGC01230 &  26.385 &  25.521 &   2.1 &   1.8 & -68 &  49.2 &   9.76 &   ---  \\ 
        UGC02936 &  60.701 &   1.966 &   2.5 &   0.7 &  30 &  51.2 &   9.85 & -19.58 \\ 
     OBC97-C04-2 & 125.872 &  21.613 &   0.4 &   0.2 &  70 &  75.2 &   8.18 & -16.69 \\ 
         F564-V3 & 135.724 &  20.076 &   0.7 &   0.5 & 156 &  10.4 &   7.11 & -12.12 \\ 
        UGC05209 & 146.268 &  32.238 &   0.9 &   0.9 &   0 &  11.0 &   7.30 &   ---  \\ 
          F568-1 & 156.526 &  22.433 &   0.2 &   0.2 &  13 &  95.5 &   9.35 & -17.49 \\ 
          F568-3 & 156.834 &  22.239 &   0.3 &   0.3 & 169 &  86.8 &   9.20 & -17.69 \\ 
        UGC05750 & 158.938 &  20.990 &   1.1 &   0.6 & 167 &  62.3 &   9.00 &   ---  \\ 
       PGC135754 & 159.365 &   2.089 &   0.6 &   0.4 &  40 & 322.0 &  10.06 & -20.99 \\ 
          F568-6 & 159.969 &  20.847 &   1.5 &   0.9 &  90 & 201.0 &  10.52 & -21.79 \\ 
         F571-V1 & 171.579 &  18.836 &   0.9 &   0.7 &  35 &  84.3 &   8.82 & -16.42 \\ 
          Malin1 & 189.247 &  14.330 &   0.3 &   0.3 &   0 & 366.0 &  10.66 & -22.90 \\ 
        PGC45080 & 195.817 &   1.469 &   0.9 &   0.2 &  84 & 178.0 &   9.99 & -18.65 \\ 
          F530-1 & 316.887 &  26.450 &   0.5 &   0.3 &  52 & 199.0 &  10.27 & -20.11 \\ 
          F533-3 & 334.305 &  25.213 &   0.9 &   0.6 & 165 & 174.0 &  10.24 & -20.44 \\ 
         NGC7589 & 349.565 &   0.261 &   1.1 &   0.7 & -60 & 120.0 &  10.01 & -21.90 \\ 
        PGC71626 & 352.635 &  -2.463 &   1.9 &   1.3 &  65 & 136.0 &  10.23 & -21.08 \\ 
\enddata 
\tablecomments{These properties were taken from the NED database  (major and minor diameters 2a and 2b, distances) and the literature (see text).}
\end{deluxetable} 

\clearpage

\begin{deluxetable}{l r r r r r r r } 
\rotate 
\tablecaption{UV properties of our LSB galaxies \label{tabuv}} 
\startdata 
        Name     &             & \multicolumn{3}{c}{------------------------ FUV ------------------------} & \multicolumn{3}{c}{------------------------ NUV ------------------------}  \\ 
                 & Last Radius &    Ap. mag    &   Asymptotic mag  & Exp time  &    Ap. mag    &   Asymptotic mag  & Exp time \\  
                 & (arcsec)    &   (AB mag)    &    (AB mag)       &  (sec)    &    (AB mag)    &    (AB mag)      &  (sec)  \\  
        UGC00568 &    5.00 & 22.81 $\pm$ 0.18 & 22.40 $\pm$ 0.17 & 3024.00  & 20.68 $\pm$ 0.06 & 18.96 $\pm$ 0.18 & 3024.00   \\
        UGC01230 &   72.20 & 16.62 $\pm$ 0.14 & 16.78 $\pm$ 0.30 &   85.00  & 16.30 $\pm$ 0.09 & 16.16 $\pm$ 0.07 &   85.00   \\
        UGC02936 &   57.10 &   ---   &      ---        &       ---            & 15.70 $\pm$ 0.09 & 15.86 $\pm$ 0.03 & 3382.45   \\
     OBC97-C04-2 &    8.50 & 20.68 $\pm$ 0.38 & 19.44 $\pm$ 0.66 &  116.00  & 19.84 $\pm$ 0.17 & 19.52 $\pm$ 0.18 &  116.00   \\
         F564-V3 &   20.30 & 19.28 $\pm$ 0.19 & 19.37 $\pm$ 0.78 &  111.00  & 19.18 $\pm$ 0.16 & 18.46 $\pm$ 0.30 &  111.00   \\
        UGC05209 &   24.00 & 18.15 $\pm$ 0.09 & 18.02 $\pm$ 0.11 &  112.00  & 17.71 $\pm$ 0.05 & 17.52 $\pm$ 0.04 &  112.00   \\
          F568-1 &   36.00 & 18.16 $\pm$ 0.03 & 18.15 $\pm$ 0.03 & 1494.00  & 17.76 $\pm$ 0.02 & 17.74 $\pm$ 0.02 & 1494.00   \\
          F568-3 &   36.00 & 17.80 $\pm$ 0.03 & 17.73 $\pm$ 0.05 & 1494.00  & 17.49 $\pm$ 0.02 & 17.41 $\pm$ 0.03 & 1494.00   \\
        UGC05750 &   44.30 & 17.47 $\pm$ 0.01 & 17.59 $\pm$ 0.01 & 3530.00  & 17.09 $\pm$ 0.01 & 17.17 $\pm$ 0.01 & 3531.00   \\
       PGC135754 &   24.50 & 19.75 $\pm$ 0.05 & 19.54 $\pm$ 0.10 & 3164.00  & 18.95 $\pm$ 0.03 & 18.70 $\pm$ 0.02 & 3164.00   \\
          F568-6 &  120.80 &   ---   &      ---        &       ---            & 16.89 $\pm$ 0.21 & 16.87 $\pm$ 0.01 & 3573.00   \\
         F571-V1 &   21.20 & 19.01 $\pm$ 0.14 & 18.47 $\pm$ 0.35 &  109.00  & 18.51 $\pm$ 0.08 & 18.14 $\pm$ 0.12 &  109.00   \\
          Malin1 &   48.00 & 19.61 $\pm$ 0.09 & 19.52 $\pm$ 0.45 & 1841.50  & 18.77 $\pm$ 0.05 & 18.30 $\pm$ 0.08 & 1841.50   \\
        PGC45080 &   48.10 &   ---   &      ---        &       ---            & 17.77 $\pm$ 0.02 & 17.50 $\pm$ 0.02 & 1377.00   \\
          F530-1 &   27.90 & 18.26 $\pm$ 0.08 & 18.12 $\pm$ 0.05 & 3392.85  & 17.53 $\pm$ 0.02 & 17.29 $\pm$ 0.01 & 9040.00   \\
          F533-3 &   49.00 & 18.11 $\pm$ 0.03 & 18.07 $\pm$ 0.04 & 4317.90  & 17.55 $\pm$ 0.01 & 17.45 $\pm$ 0.01 & 5924.05   \\
         NGC7589 &   38.30 & 17.70 $\pm$ 0.03 & 17.53 $\pm$ 0.02 & 1495.00  & 17.21 $\pm$ 0.01 & 17.05 $\pm$ 0.02 & 1495.00   \\
        PGC71626 &   84.40 & 17.05 $\pm$ 0.09 & 16.96 $\pm$ 0.03 & 3411.00  & 16.33 $\pm$ 0.01 & 16.21 $\pm$ 0.05 & 3411.00   \\
\enddata 
\tablecomments{UV photometry: last radius measured, aperture magnitude within the last radius, asymptotic magnitude and exposure time for the FUV and NUV bands of GALEX \citep[see][ for details on these values and how they are obtained]{gilatlas}}

\end{deluxetable} 

\clearpage

\subsection{UV observations and photometry}

We have NUV imaging data for all galaxies.
FUV data is available for all but three objects.
Five LSB galaxies were observed in the framework of the
shallow GALEX All sky Imaging Survey ($\sim$ 100 seconds of exposure
time). We have deeper images for the other galaxies, with exposure
times ranging between $\sim$ 1500 and 9000 seconds depending on the
survey/program (Nearby Galaxy Survey, Guest Investigator program). The
exposure time for each observation is given in Table \ref{tabuv}.
The GALEX images can be seen in the Fig$.$ \ref{figindividual} of the
Appendix.

GALEX photometry was performed using the same code as for the GALEX
Ultraviolet Atlas of Nearby Galaxies \citep{gilatlas}, and our results
are given in Table \ref{tabuv}. For each of the two GALEX bands, we
provide the magnitude measured within the last isophote that could be
measured in FUV\footnote{The radius of this isophote 
corresponds to the position where the error 
in the azimuthally-averaged surface brightness becomes larger than 0.8 mag.}
('Aperture Magnitude' in Table  \ref{tabuv}), 
and an asymptotic magnitude, obtained by extrapolation of
the curve of growth \citep[see details in ][]{gilatlas}. In a few cases,
the uncertainties were too large to actually perform this extrapolation
(FUV magnitude of UGC00568 and F564-V3 and the NUV magnitude of OBC97-C04-2)
and we are giving instead the magnitude measured in the largest possible 
radius.
Due to the faint nature of our objects, the asymptotic magnitudes suffer
relatively large errors. Unless stated otherwise, we thus use the aperture 
magnitudes.

\subsection{Ancillary data}

Thirteen of our eighteen galaxies have been covered by the Sloan
Digital Sky Survey (SDSS, DR5). For these, we downloaded from the SDSS
skyserver\footnote{http://cas.sdss.org/astro/en/} the images in the
five SDSS bands, and performed the same surface photometry as on the
GALEX images.  In Fig.~\ref{figindividual}, we show the $g$ band image
($g$ and $r$ images are reasonably deep, but our objects are faint in
other bands).  We computed the integrated magnitude within the last
radius for which the FUV flux was measured, or NUV for galaxies
without FUV data (``Last Radius'' in Table \ref{tabuv}).
The SDSS images are rather shallow, especially for LSB galaxies. As a
result, our integrated magnitudes have quite large error-bars. On the
other hand, we computed them in a similar way as the GALEX ones, and
within the same apertures.

We also included in Fig. \ref{figindividual} a few other magnitudes at
various wavelengths: first, the 2MASS J,H, and K total magnitudes, as
given in the NED database for 10 galaxies.  Our study overlaps with a
few works on LSB galaxies, from which we also took total magnitudes,
as published : \citet{deblok95} for F564-V3, F568-1, F568-3, F571-V1,
UGC~01230, UGC~05209, UGC~05750; \citet{mcgaugh94b} for UGC~01230, F568-6; 
and \citet{hunter06} for F564-V3, UGC05209.
While we performed the SDSS photometry following the same procedures
as for the UV, these other studies are independent, and some
differences might exist in e.g. the position, Position Angle,
aperture. Nevertheless they give an idea of the overall shape of the
galaxy SED when compared to other magnitudes as is done in
Fig$.$\ref{figindividual}.

\subsection{Reference samples}

In order to compare our results to high surface brightness galaxies,
we considered two large samples typical of ``normal'' (non-LSB) 
star-forming galaxies.

The first one is the GALEX Atlas of nearby galaxies \citep{gilatlas}.
The big advantage of this sample is that the NUV and FUV photometry
were performed in the same way and with the same code as for our
galaxies. Galaxies in the Atlas are not selected on the basis of their
surface brightness, but are representative of nearby galaxies.
Their properties are actually consistent with the Nearby Field Galaxy
Survey of \citet{jansen00}.  Although some LSB (or
intermediate-surface brightness) galaxies might be included, the
typical star-forming galaxies in the Atlas are high
surface-brightness objects.  For this sample, we queried the LEDA
database \citep{paturel03} to obtain HI magnitudes and convert them
into HI masses.

The second sample is a set of star-forming galaxies from the GOLD Mine
database \citep{gavazzi03}, including multi-wavelength data for a
large number of cluster galaxies.  The advantage of this sample is
that it includes a determination of the HI deficiency. The HI
deficiency is defined as the logarithmic difference between the
average HI mass of a reference sample of isolated galaxies of similar
type and linear dimension and the HI mass actually observed in
individual objects \citep{gavazzi05}.  We excluded all galaxies with
HI deficiencies larger than 0.3, what is typically found in perturbed
galaxies (e.g. ram-pressure stripping within clusters).
For this sample, the attenuation in the UV was already
estimated \citep[following the method of][]{boselli03}, 
so the FUV and NUV magnitudes can be corrected for internal
extinction.

\section{Spatial extent of the UV emission}

\subsection{Extended emission}


We inspected our images to determine which of our galaxies 
present a XUV-disk like morphology, except for the five galaxies from the
All sky Imaging Survey (AIS) which are too shallow to really discuss
this point.

\citet{thilkerxuv} defined two types of XUV-disk galaxies.
Type 1 XUV-disk galaxies present structured UV-bright emission complexes 
beyond the anticipated location of the star formation threshold 
(corresponding to a NUV surface brightness of about 27.35 
AB mag /arsec$^2$). Six of our galaxies correspond to this case:
NGC~7589, PGC~135754, PGC~71626, F~533-3, Malin~1, F~568-6.
Type 2 XUV-disk galaxies are defined by \citet{thilkerxuv} 
as galaxies forming stars (i.e. UV bright) over an area much
larger than the spatial extent of their old stellar population
(as traced by near infrared light). Although we do not have 
deep enough K-band images to use their quantitative definition,
the comparison of UV and optical images strongly suggests that F~568-1,
F~530-1 and F~568-3 are actually type 2 XUV-disk galaxies 
\citep[we thus find the same fraction of type 1 vs type 2 as ][2 for 1, even if we have a very small number of objects]{thilkerxuv}.

\citet{thilkerxuv} found that about 30 \% of 
the galaxies in the GALEX Atlas of nearby galaxies fall into any of 
the XUV categories, and \citet{zaritsky07} found at the 90\% confidence level
that 27 \% of the spirals have UV sources in their disks at radius 
between 1.25 and 2 optical radii, showing that extended UV emission
is common in nearby spirals. 
Out of 13 deep UV images of LSB galaxies, 
nine present clear signs of extended emission, thus this phenomenon is
even more frequent in LSB than normal galaxies, although we might be 
suffering from poor statistics.

In the following, we will discuss the UV light distribution in 
a few galaxies of special interest.
%
Recent studies suggested that Malin~1 is an early-type galaxy
surrounded by a huge LSB disk \citep{sancisi07,barth07}. The UV images
of Malin~1 show the central part of the galaxy very clearly. In the
outer disk, we observe several diffuse emission regions, probably
corresponding to relatively recent star formation within the LSB disk
\citep[see also ][]{thilkerxuv}.
%
Malin~2 (F568-6) presents a spectacular extended UV disk with a clear
spiral morphology, while it can only be guessed in optical (DSS or
SDSS images).
Spiral patterns at large radii are observed in several others
of our galaxies (PGC~71626, NGC~7589, F530-3). 
They are often barely visible in the optical, 
but the contrast is much more favorable
at the UV wavelengths.
This is due to the fact that the arm-interarm contrast is much more
favorable in the UV if star formation is enhanced in the arms (UV
emitting stars have short lifetime, thus are found closer to their
formation locus), and owing to the low background in the UV.
Some other galaxies may harbor star formation within
spiral patterns that cannot be seen in our images because of short 
exposure times or low spatial resolution (F568-3, F568-1).

\subsection{Optical to UV size ratio}

\begin{figure}
\includegraphics[width=8cm]{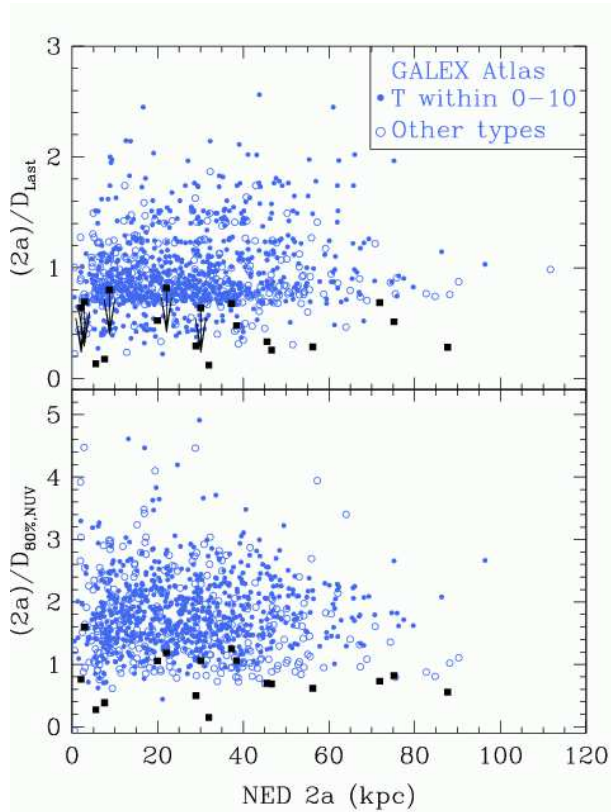}
\caption{Ratio of the optical to the total NUV diameter (top) and 80
  \% (bottom) NUV diameter (diameter including 80 \% of the light) as
  a function of the optical diameter of the galaxies in our sample
  (squares). Open and filled circles are the same quantities for the
  GALEX Atlas of Nearby Galaxies \citep{gilatlas}. The last radius depending
on the exposure time, galaxies from the All sky Imaging Survey would have
larger UV diameter (and lower optical to UV diameter ratio) if observed at the depth of the GALEX Atlas. In that
case, the ratio would be smaller than the one we determined, 
what we indicated in the top panel with arrows. 
\label{figradext}}
\end{figure}

Fig. \ref{figradext} shows that while our LSB galaxies cover the range of
optical diameters found in the GALEX Atlas of Nearby Galaxies \citep{gilatlas}, 
the optical to UV diameter ratio in LSB galaxies is on average
much smaller (by about a factor 2) than the same quantity for the
galaxies in the Atlas, considering either the last UV isophote (top) or the 
diameter in which 80 \% of the UV flux is contained. 
Thus, the UV light in LSB galaxies is on average more
extended with respect to the optical than ``normal'' galaxies.

The diameter corresponding to the last radius where the UV isophote
was computed (top panel of Fig.~\ref{figradext}) corresponds to a limiting signal to
noise level, sensitive to the depth of the images. The comparison
to the one measured in the GALEX Atlas of nearby galaxies makes sense
for most galaxies since the exposure time are of the same order of magnitudes
(within a factor of a few).
We can thus make a quantitative comparison of this diameter ratio in
our LSB galaxies and in the GALEX Atlas.
This is not the case however for the data taken from
the All sky Imaging Survey. For them, if we had deeper images (similar to the
one in the Atlas), we could  measure UV isophotes further away from the center 
than we actually did, and we would find lower optical to UV ratio. This effect,
indicated by arrows in Fig.~\ref{figradext}, can thus only strengthen our conclusion.

The fact that the optical/UV diameter ratio of LSB galaxies is low may be a
sign that these galaxies are relatively un-evolved objects, having
formed stars in the past only in their central part, with large-scale
star formation occurring currently in a large HI disk. This is
consistent with the analysis of \citet{bell00} who found optical-near
infrared color gradients showing younger ages in the outer parts of LSB 
galaxies.

\section{Star Formation Rates and Star Formation Efficiency in LSB galaxies}

\label{secsfr}

\subsection{Theoretical expectations}

The models for LSB and HSB galaxies of
\citet{boissier03} and \citet{boissier00} predict that the the lower
surface density and larger radial extent in massive LSB objects 
make them less efficient at forming stars by a factor 5 to 10. 
These models are of the general type that consider LSB galaxies as
analogues to HSB galaxies, but with larger specific angular momentum
\citep{jimenez98,dalcanton97}.  For the same total mass, the surface
densities are lower. If the star formation density is not a linear
function of the gas density, a lower star formation efficiency is to
be expected. It is the case of the models of \citet{boissier03} where
the star formation density ($\Sigma_{SFR}$) is proportional to
$\Sigma_{GAS}$ to a power 1.5, and to the inverse of the radius. It is
also the case of the models of \citet{jimenez98} who assume
$\Sigma_{SFR} \propto \Sigma_{GAS}^{1.5} \Sigma_{STARS}^{0.5}$. Indeed,
in both cases the ratio $\Sigma_{SFR}/\Sigma_{GAS}$ will be smaller for larger,
lower densities LSB galaxies. 
%
%
\citet{deblok96} and \citet{mcgaughthesis} showed that the gas surface
densities in LSB are down by a smaller fraction than the stellar surface densities
when comparing to HSB galaxies (a factor of 5 or 10 in surface brightness
corresponding to only 2 to 3 in gas densities). 
This result is qualitatively consistent with such star formation law:
for instance, considering the star formation law used
by \citet{boissier03}, we have:
\begin{equation}
\label{eqcontrast}
\frac{\Sigma_{SFR,HSB}}{\Sigma_{SSFR,LSB}}=\left( \frac{\Sigma_{GAS,HSB}}{\Sigma_{GAS,LSB}} \right )^{1.5}  \frac{R_{LSB}}{R_{HSB}}
\end{equation}
where $R_{LSB}$ and $R_{HSB}$ are typical sizes of LSB and HSB
galaxies.  Assuming they encompass the same total HI mass, we can
write $R_{LSB(HSB)} \propto \Sigma_{GAS,LSB(HSB)}^{-0.5} $, and then
rewrite equation \ref{eqcontrast} as
$(\Sigma_{SFR,HSB}/\Sigma_{SFR,LSB})$ $=$ \\
$(\Sigma_{GAS,HSB}/\Sigma_{GAS,LSB} )^{1.5+0.5}$.  With this simple
back-of-the-envelope calculation, we find that a factor 3 difference in the
gas density can produce a factor 9 difference in the star formation rate surface
density between LSB and HSB galaxies. If the gas surface density is
approximatively constant during the history of the galaxy, the same
factor applies to stellar surface densities.

Note that observationally, we use integrated 
values rather than surface densities. 
It is equivalent because the UV diameter (in which the UV is detected)
is roughly the diameter where the gas column density reaches 1 \msun{}
pc$^{-2}$, at least for those galaxies for which it has been measured
\citep[HI diameters are however quite uncertain;][and thus
the gas surface densities defined in that way, and the 
the star formation law derived from them are crudely
defined]{matthews01}.

\citet{gerritsen99} performed N-Body simulations in which star
formation and feedback are implemented in a very different way (based
on the Jeans mass).  In their case, they did not find differences in
the SFR with respect to the HSB case; however they still found lower
SFR for LSB galaxies when including the effects of the low metallicity
in LSB galaxies, lowering the cooling efficiency and thus the amount of 
star formation. Based on their
work also, a lower efficiency should then be expected in LSB galaxies, but
due to their lower metal content.

The lower global efficiency to form stars in LSB galaxies was also expected
in the models of \citet{vanderhulst93} because their gas surface 
density is smaller than the ``star formation threshold''.

\subsection{Empirical determinations of the SFR}

In this section, we assume the standard conversion factors from the UV
to derive Star Formation Rate following \citet{kennicutt98}.  The
results are given in Table \ref{tabsfr}.  We caution that the these
conversions assume that the star formation rate is roughly constant
over a few 10$^8$ yr, that there is no attenuation by dust (or this
effect has been corrected), and that the galaxy Initial Mass Function
is standard (close to Salpeter).  We will see in the next
sections that some of these assumptions might not be valid, what
will greatly affect our result.

\begin{figure}[h]
\includegraphics[width=8cm]{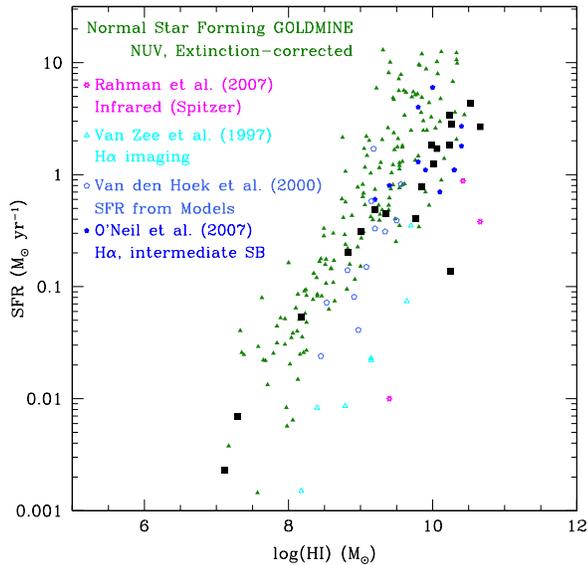}
\caption{NUV-derived SFR as a function of the HI mass in LSB galaxies (filled squares), compared to
  extinction corrected SFR in normal star forming galaxies (from
  GOLD Mine), and other determinations in LSB galaxies.\label{figsfr}
These values are however very uncertain
}
\end{figure}

Our UV SFRs are larger than the ones determined in various papers
concerning LSB galaxies (see Table \ref{tabsfr} and Fig \ref{figsfr}).  There
are several reasons for this that we discuss below: i) in the absence
(or deficiency) of dust (as it seems to be the case, see next section), 
the infrared will provide under-estimated
measurements of the star formation rate \citep{rahman07}.  
ii) Some studies \citep{burkholder01} provide nuclear SFR while we
provide integrated measurements for these significantly extended
galaxies.
iii) The SFR of \citet{vanzee97} are a bit smaller than ours 
for the same HI mass (the same is true for a given B magnitude).
They are derived from H$\alpha$ data. This could indicate an age (the
most massive stars have disappeared) or an IMF 
(massive stars were not formed in the first place)
effect. Note that their sample concern mostly dwarf galaxies.
\citet{oneil07} also determined SFR from H$\alpha$, but for more
massive galaxies, with intermediate surface brightness. Their results
are quite similar to ours (see Fig. \ref{figsfr}).

On the other hand, the SFRs obtained by \citet{vandenhoek00} from
modeling broadband and HI content are quite in agreement with the
ones in our sample at the same magnitude, or HI mass (one of
their galaxy has a large SFR, but is uncertain due to contamination by
other sources). Notice that they sample relatively faint LSB galaxies
and not the more massive ones for which we have more data and see more
differences with respect to ``normal'' spirals).

In Fig. \ref{figsfr}, we compare the relation between SFR (derived
from NUV) and HI in our sample of LSB galaxies to the one in nearby
star forming galaxies from GOLD Mine, corrected for extinction
effects. We find that the SFR in the massive LSB is lower, thus the
efficiency of forming stars (or at least NUV emitting stars) is lower
in such galaxies (we checked that the massive GOLD Mine galaxies have still
larger SFR if we use data uncorrected for extinction for them). This 
is in nice agreement with the theoretical expectations discussed above
that the star formation efficiency should be lower (see however the next sections in which 
we show that determining SFR in LSB galaxies may be a harder task than 
what is done here and in most empirical works).
The few low-mass LSB galaxies we have in our sample seem 
similar to ``normal''  star forming galaxies for their mass,
in terms of star formation rate.

\clearpage

\begin{deluxetable}{p{3.5cm} r r r p{5cm} } \tablecaption{SFR deduced
    from the \citet{kennicutt98} conversion factor, compared to SFR
    from other studies (caution should be taken concerning these
    values, see section \ref{secsfr})\label{tabsfr}} 
\startdata
  Name     &  FUV SFR  & NUV SFR &  Other & Ref for other, comments  \\
  &  \sfru{}   &  \sfru{}   & \sfru{} & \\
  UGC00568 &   0.0165   &  0.1174 & & Early-type SED, measured only in the central 5 arcsec \\
  UGC01230 &   0.3311   &  0.4446 &  &\\
  UGC02936 &  ---       &  0.8366 & \\
  OBC97-C04-2 &  0.0184 & 0.0399  &\\
  F564-V3  & 0.0013     & 0.0014  &\\
  UGC05209 & 0.0040     & 0.0061  & \\
  F568-1   &  0.3020    & 0.4365  &   0.31 & \citet{vandenhoek00}, model\\
  F568-3   &  0.3476    & 0.4624  &   0.33 & \citet{vandenhoek00}, model  \\
  UGC05750 &  0.2426    & 0.3443  &\\
  PGC135754 &  0.7938   & 1.6585  &\\
  F568-6   & ---        & 4.3092  &\\
  F571-V1  &  0.1076    & 0.1705  & 0.14  & \citet{vandenhoek00}, model \\
  Malin1   &  1.1667    & 2.5291  & 0.38 &  \citet{rahman07}, infrared\\
  PGC45080 & ---        & 1.5026  &\\
  F530-1   & 1.1959     & 2.3426  &\\
  F533-3   & 1.0498     & 1.7583  &\\
  NGC7589  &  0.7284    & 1.1438  &\\
  PGC71626 &  1.7025    & 3.3043  &\\
  \hline
  LSB galaxies {\scriptsize ($-15.14$ $>M_B>$ $-21.17$)} \vspace*{0.2cm}  & & & $\sim$ 0.2  & \citet{mcgaugh94b}, model \\ 
  dwarf LSBs            \vspace*{0.2cm}         &          &         & 0.0083-0.35 & \citet{vanzee97}, H$\alpha$ imaging\\
  sub\-sample of APM, $\mu_0>21$\vspace*{0.2cm} &          &         &  0.17 $\pm$ 0.36      & \citet{burkholder01}, H$\alpha$, nuclear \\
  16 rela\-ti\-ve\-ly faint LSBs  \vspace*{0.2cm}   &         &          & median 0.15 & \citet{vandenhoek00}, models \\
  intermediate surface brightness               &          &         &   0.3-5    &\citet{oneil07} \\
\enddata
\end{deluxetable}

\clearpage

\section{FUV-NUV color of LSB galaxies}

\begin{figure}
\includegraphics[width=8cm]{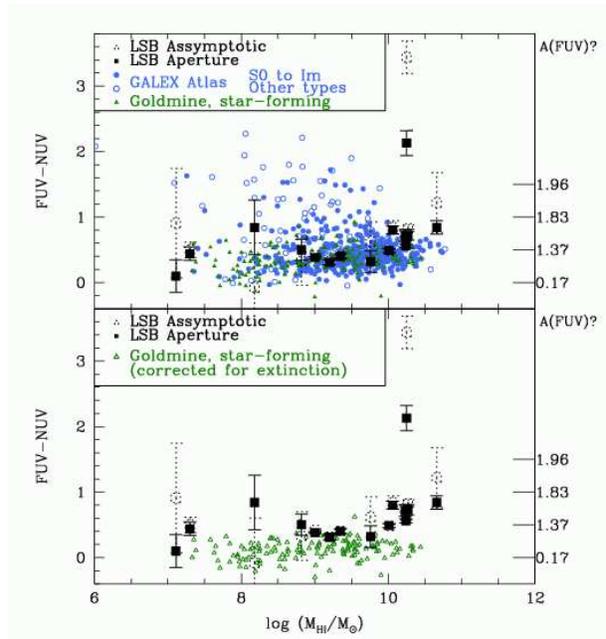}
\caption{Observed FUV-NUV color of our sample of LSB galaxies (using
  aperture, squares, or asymptotic, open circles, magnitudes) as a
  function of the HI mass, compared to a sample of normal star forming
  galaxies from GOLD Mine, and to the GALEX Atlas of Nearby Galaxies.
  In the top panel, the LSB color is compared to the FUV-NUV color
  observed in various samples. In the bottom panel they are compared
  to the FUV-NUV color corrected for extinction in the GOLD Mine
  sample.  The right axis indicates the amount of extinction derived
  using the relationship between the infrared to UV ratio and FUV-NUV
  in \citet{boissier06}, and the A(FUV) vs infrared to UV ratio in
  \citet{buat05}.  These fits are only valid for FUV-NUV
  $\lesssim$~1.5. We included a question mark in the label A(FUV) to
  stress that this is the extinction one would derive from the color
  for usual galaxies. However, this conversion may not be valid for
  LSB galaxies (see text of section 5).
\label{figcolor}}
\end{figure}

\subsection{Observations}

We checked that the FUV-NUV color profiles of our galaxies (not shown) are
quite flat, so that the integrated color is similar to the color all over
the disk, including the outer regions. UGC~00568 is an exception: this
galaxy is barely visible in the FUV image, and it was not possible to
extract a reliable profile in this band. 
As a result, the FUV-NUV color corresponds only to the central 7
arcsec. The rest of the disk is probably also red since it is not
detected in the FUV band.  This is the only object in our sample that
shows a SED similar to those of early-type galaxies (Fig.
\ref{figindividual}).
Note that even if the magnitudes are integrated within a larger
radius, the NUV-2MASS colors are still consistent with an early-type
galaxy SED.  This is our only case of extremely red LSB despite a huge
HI reservoir.

In normal star forming galaxies, the total infrared (TIR) to UV
emission ratio of a galaxy is a good proxy for the attenuation 
\citep{buat96,gordon00,panuzzo03}. 
Several studies \citep{boissier06,gilatlas,cortese06,seibert05} have
shown that a relation exists between this ratio and the FUV-NUV color
in star forming galaxies, even if it is shifted (towards lower
extinction for the same color) with respect to the classical
relationship found with IUE in starbursts
\citep[e.g.][]{heckman95,meurer95,meurer99}.
Thus, it is expected that star forming galaxies with low extinction
should have blue FUV-NUV colors, close to zero, while redder colors
should indicate some amount of extinction. 

The FUV-NUV colors of our LSB galaxies are similar to the ones
measured in normal star forming galaxies, and follow similar trends
with various quantities (we show in Fig. \ref{figcolor} the FUV-NUV
color vs the HI mass). From this figure, it even seems that for the
same mass, the color is marginally  on the red side of normal galaxies, especially
for LSB galaxies with HI masses above 10$^{10}$ \msun.
%
The bottom panel of Fig. \ref{figcolor} clearly shows that LSB are
almost all redder than the FUV-NUV color of star forming galaxies
when they are corrected for attenuation. 
Such corrections are uncertain, however the corrected colors do
correspond to what is expected in the stellar populations of star
forming galaxies (FUV-NUV $\sim$ 0, see section 5.3).
A natural explanation of the
FUV-NUV color of LSB galaxies is thus that they suffer large internal
attenuation, similarly to HSB normal star forming galaxies.
We combined the \citet{buat05} relationship (FUV attenuation as a
function of the TIR/FUV ratio), and the \citet{boissier06}
relationship (TIR/FUV as a function of FUV-NUV in nearby spirals) in
order to convert the FUV-NUV color into the corresponding amount of extinction
in normal galaxies. An extinction scale computed in that
way is shown in the right axis of Fig. \ref{figcolor} (we cannot use
directly the TIR/FUV ratio since we do not have infrared measurements
for the galaxies in our sample, with the exception of Malin~1, see section
\ref{secext}). This gives us the amount of attenuation A(FUV) necessary
to redden the FUV-NUV color to the observed level, about 1.5 mag
for the massive LSB galaxies.
We study this assumption in section \ref{secext}.

We emphasize that this calibration of the attenuation with the
FUV-NUV ratio should be valid only for star forming galaxies: if star
formation was quenched some time ago (early type, truncated SFR), this
relation shall fail \citep[e.g.][]{boselli06,cortese08} because
old stars would have a red FUV-NUV color even in the absence of dust.
However our SEDs are consistent with those of Irregular, and late-type spirals, 
in which such quenching should not a priori occurs, except for UGC~00568.
A few other galaxies have not so late type (similar colors to the Sbc
template), but they are not systematically the redder ones in FUV-NUV.
We note also that optical/red images show a more concentrated
morphology than the UV, i.e. the color of star forming regions,
especially outer ones are bluer than the integrated one (for the
galaxies with SDSS data, we verified this point with the NUV-$r$ color
profile). This is an indication that the FUV-NUV and e.g NUV-$r$
color do not trace the same stellar population.
In this case the FUV-NUV red color could correspond to 
a star formation that was quenched recently in the young regions,
while the optical colors would be more sensitive to a smooth
star formation history on the timescales corresponding to 
older populations.
Alternatively, redder FUV-NUV colors could also be explained by
an IMF effect (we will come back to such scenarios in section \ref{secage}).

\subsection{Are LSB galaxies affected by dust attenuation ?}

\label{secext}

As discussed above, the FUV-NUV color in our LSB galaxies could indicate 
significant amount of attenuation, increasing with the HI mass of galaxies.
However, it is generally believed that LSB disks are deficient
in dust with respect to their high surface brightness counterparts
\citep[e.g.][ and references therein]{rahman07}, based of their
generally blue colors, low densities, low metallicities and 
deficiency in molecular gas. Actually, recent
studies show that LSB galaxies do contain molecular gas (although in
smaller amount with respect to HSB disks) that, being  localized in
isolated regions, is difficult to detect \citep{oneil04,das06}.  
We should also note that measuring the CO molecular emission and
converting it to gas masses is quite uncertain due to the uncertainty
on the conversion factor from CO to H$_2$ \citep{boselli02}, 
especially at the low metal abundances and densities found in these galaxies.
Among the reasons letting to think that the attenuation in LSB galaxies is weak,
we should also note that LSB disks are found to be transparent by
\citet{holwerda05}, based on the count of distant field galaxies seen
through disks.


\begin{figure*}
\includegraphics[width=8cm]{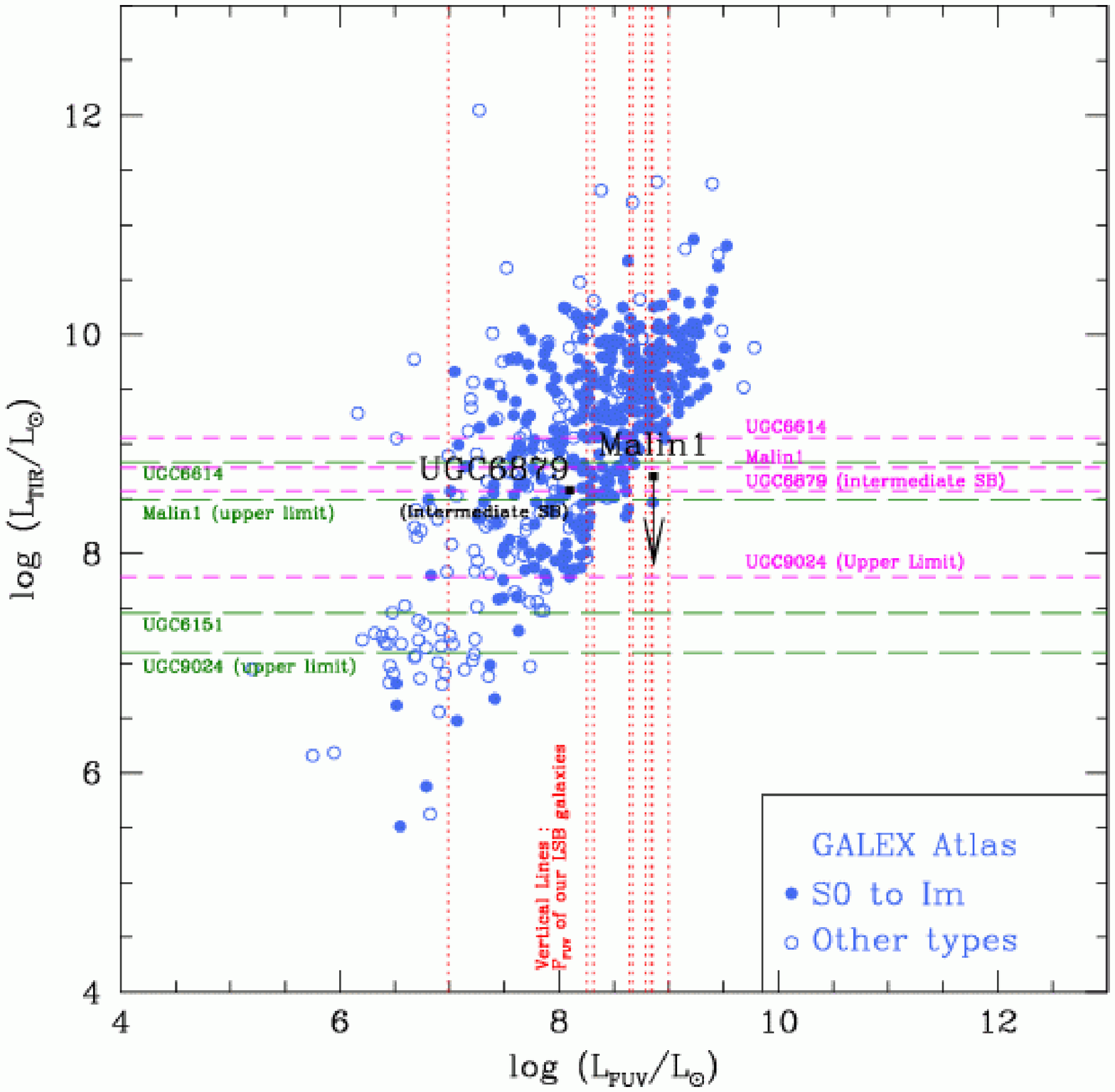}  \includegraphics[width=8cm]{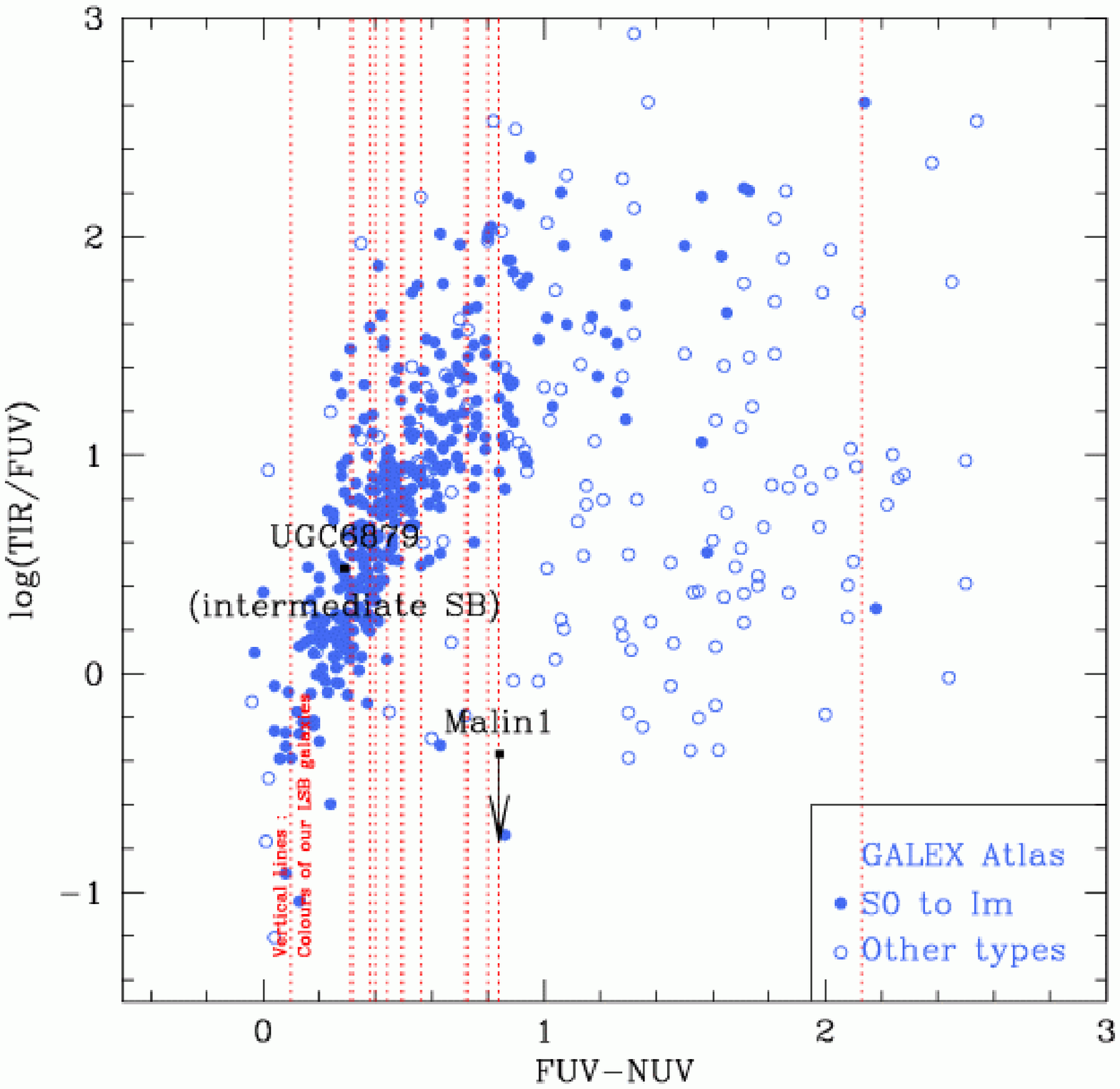} 
\caption{Left : FUV vs TIR luminosities. The horizontal (vertical) lines indicate TIR (FUV)
values for galaxies in \citet{rahman07} and \citet{hinz07} (our work, only showing galaxies in the same
$M_B$ magnitude range as these studies).
Right : TIR to FUV ratio vs FUV-NUV color.
\label{figdust}}
\end{figure*}

\begin{figure}
\includegraphics[width=8cm]{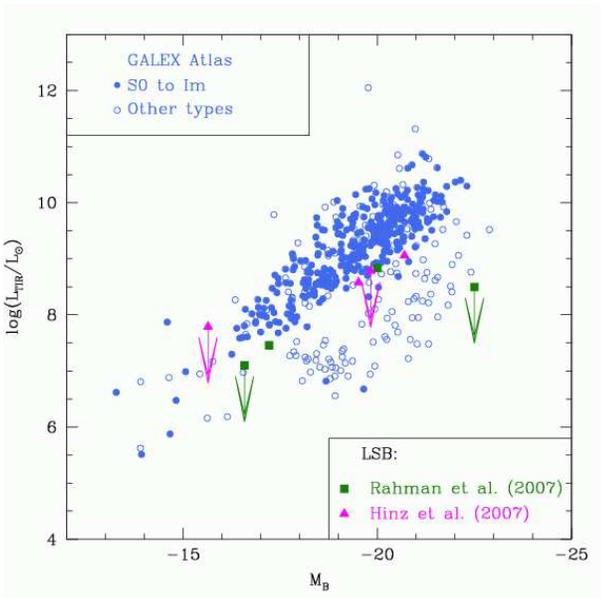}  
\caption{TIR luminosities as a function of the blue band absolute magnitude for
galaxies of the GALEX Atlas of nearby galaxies (circles) and for LSB galaxies 
(squares and triangles).
\label{figdust2}}
\end{figure}

Very recently, it became possible to study the far infrared dust emission 
in LSB galaxies owing to the \emph{Spitzer} Space Telescope.
\citet{hinz07} reported observations of 5 LSB galaxies with Spitzer, three
of them also analyzed (with slightly different results) in \citet{rahman07}, 
together with one ``intermediate'' surface brightness galaxy. \citet{hinz07}
concluded that the far-infrared emission is weak when compared to normal
star forming galaxies and that LSB galaxies contain less and/or
colder dust. They suggested that the dust is detected in the
infrared for galaxies with large amounts of star formation.
From the MIPS data in these papers, we computed the total infrared
(TIR) emission \citep[using Eq. 4 of][]{dale02} and show them as
horizontal dashed lines in the left panel of Fig$.$ \ref{figdust}.
The vertical dotted lines indicates FUV fluxes for LSB galaxies in our
sample with blue band magnitudes in the same range (-23$<M_B<$-17) as
the galaxies in \citet{rahman07} and \citet{hinz07}.  Assuming LSB
galaxies with similar $M_B$ are indeed similar, LSB galaxies should
lie in this diagram in the region where horizontal and vertical lines
cross (notice that many horizontal lines are actually upper limits).
The TIR emission of the galaxies in the GALEX Atlas of Nearby galaxies \citep[computed
from IRAS 60 and 100 $\mu$m fluxes as in ][]{dale01} is also shown.  
Measured TIR values in massive LSB galaxies are clearly weaker that
the values found in normal star forming galaxies with the same FUV
flux.  Malin~1 is the only true LSB galaxies for which we have both
TIR and FUV values, and that can be definitively placed in this
diagram. We find that its TIR flux (actually an upper limit) is
smaller than for any normal star forming galaxies with the same FUV
flux.  We note that UGC~06879 is considered by \citet{rahman07} as an
intermediate surface brightness galaxy.  Using the UV fluxes from
\citet{gilatlas}, we see that it is located among normal star forming
galaxies in this diagram.

The right part of Fig. \ref{figdust} shows the ratio TIR/FUV vs the
FUV-NUV color.  We can see the trend followed by normal star forming
galaxies mentioned above: redder galaxies are more extinguished. The
intermediate surface brightness galaxy from \citet{rahman07} is on
this trend.  The upper limit on the far-infrared emission of Malin~1
from \citet{rahman07} corresponds to a very low upper limit on this
ratio, resulting in an upper limit of 0.4 mag of attenuation with the
calibration of \citet{buat05}.  \citet{hinz07} obtained lower numbers
for the infrared fluxes upper limits of Malin~1, stressing the
difficulty to determine them in LSB galaxies. Adopting their results,
we obtain an even lower value for the maximal FUV attenuation in
Malin~1 : 0.1 mag \citep[similar attenuations of 0.39 and 0.2 mag are
obtained using the calibration of ] [based on the FUV-g color of
Malin~1]{cortese08}. In any case, this is much lower than the
attenuation necessary to redden the FUV-NUV color estimated in section
5.1 ($\sim$ 1.5 mag).

Unfortunately, at the present, we have both FUV and TIR measurements
only for Malin 1. Instead of comparing the TIR and FUV emission of
galaxies with similar blue band magnitude $M_B$, as we did above, we
can also directly compare the relationship between the TIR emission
and $M_B$ for the few LSB galaxies for which we have this information
to the one obtained with the GALEX Atlas of nearby galaxies
(Fig$.$\ref{figdust2}).  Here again, we see that LSB galaxies have
systematically low dust emission in the infrared with respect to
normal star forming galaxies with the same $M_B$. This suggests they
suffer lower amount of extinction.

Putting together the results from this section, although the trends
followed by the FUV-NUV color in LSB galaxies are similar to the ones 
observed in HSB galaxies, and for which it is due (we believe) to an
attenuation effect, other considerations (especially infrared fluxes)
clearly suggest that there is very little extinction even in the more
massive, redder LSB galaxies. In the following, we seek for other
explanation for their FUV-NUV color.

\subsection{Age or IMF effect ?}

\label{secage}

Fig. \ref{figdust} shows that Malin~1 (and probably other massive LSB
disks) have a low value of  TIR/FUV for a large value of FUV-NUV, with respect
to ``normal'' (non-LSB) star forming galaxies. 
\citet{kong04}, for instance, used models to show that the position of a
galaxy in this diagram may depend on its star formation history.
In order to match the position of Malin~1 with their models, it would be necessary to
assume extremely low current to past average star
formation rate (birthrate $b$ parameter). This is however not very
compatible with the idea that LSB galaxies are ``young'', i.e.
to the light of their
stellar population and chemical state (large gas fraction, low
metallicity, blue optical colors in many galaxies...).  The spectrum
resulting from such history should be similar to early-type galaxies.
Only one galaxy in our sample has such a spectrum (UGC~00568). The remaining of 
them are indeed similar to star forming Irregular and late spirals
types. One possible explanation would be that star formation proceed by bursts with
quiescent phases longer than the UV emission timescale (a few 100 million years),
but shorter than the optical emission timescale (Gyr scale).
This could produce during quiescent phases red FUV-NUV colors, relatively blue optical colors,
and low birthrate parameters (UGC~00568 would be an extreme case in which the quiescent phase
was long enough to also affect optical colors).

Indeed, in the absence of extinction, the FUV-NUV color
can be a good indicator of the age of a star cluster
\citep{bianchi05}.  FUV-NUV color ranging from about -0.2 to 1 would
correspond roughly to ages of the clusters within 3 Myr to 360 Myr
according to this paper.  
If the FUV-NUV color results from such an age effect, it still remains
to be explained its dependence on the B-band magnitude or HI mass
(redder colors are found in the most massive LSB disks), since it is
not obvious why the age of the most recent star-formation event should
behave in this way. Part of the trend could actually be due to a metallicity
effect since more massive galaxies are more metal-rich (see section 6
for a detailed derivation of this age, taking into account the metallicity). 

An alternative would be that the Initial Mass Function (IMF) is
different in low density regions, favoring lower mass stars which are
redder.  Indeed, \citet{weidner05} found that the ``Integrated
Galaxial'' IMF is steeper than the universal IMF, assuming stars are
born in clusters following a mass function whose the maximum mass
($M_{ecl,max}$) is a function of the galaxy SFR (stars are born
following the IMF within these clusters). They predict a strong effect
at low star formation rates, and predict that stars more massive than
25 \msun{} will never form in low mass LSB galaxies. When the IMF gets
steeper, favoring less-massive stars, we should obtain redder colors.
Moreover, with such an effect, we should obtain that less massive
galaxies, in which lower amount of star formation takes place should be
redder, while we observe the opposite trends (see Fig. \ref{figcolor}
for FUV-NUV vs the HI mass, a similar trend is found with the
magnitudes in either B, NUV or FUV bands).
A steep IMF in LSB galaxies is also suggested by \citet{lee04} who found
it provides a better agreement with observations of mass to light ratios
and optical colors (B-R,B-V,B-I). However their result is based on a
single burst scenario (a few Gyr old). 
If the star formation history is more complex, for instance composed
of a few Gyr old burst on top of an older underlying stellar
population, a standard (stellar) IMF could be accommodated since the
single burst scenario would require an excess of low-mass stars to
compensate for the contribution of the most evolved stellar
population.
Until more complex star formation histories are
considered in similar studies, conclusions concerning the IMF should
be considered cautiously.
Another constraint on the IMF comes from the fact that LSB galaxies
follow the Tully-Fisher relationship \citep{mcgaugh95}, what would
be hard to understand if their IMF was extremely different than the one
in HSB galaxies.

\label{secmodelsam}

\begin{figure}[t]
\includegraphics[width=8cm]{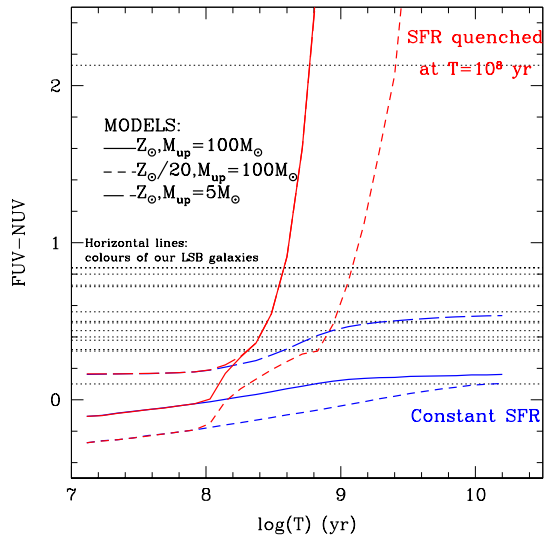}
\caption{Effect of the age or IMF on the FUV-NUV color. 
Horizontal lines correspond to the values measured in our sample, the various
curves to different models as indicated in the figure.\label{figimfage}}
\end{figure}

In order to test the effect of age and IMF on the FUV-NUV color in a
very simple way, we used the code of chemical and spectrophotometric
galactic evolution of \citet{boissier99} to compute the evolution
of this color in two scenarios~: a constant star formation rate, and a
post-burst scenario, in which we use a constant star formation rate
for 10$^8$ yr, 
time after which the star formation rate is quenched. We assume that
the FUV and NUV luminosities are then dominated by the fading of the
stars created during this event. For each of these scenarios, we made
three assumptions: adopting a solar metallicity and the
\citet{kroupa93} IMF, a low metallicity (a 20th of solar) with the
same IMF, and finally a solar metallicity and the \citet{kroupa93}
IMF but truncated at a very low mass of 5 \msun{} \citep[in order to
test if the IMF of LSB galaxies is truncated at high masses, as suggested by ][]{weidner05}. 
We note that in the case of the post-burst scenario, our results
slightly depend on the quenching time used in our computation (10$^8$
yr).  However, the differences obtained by varying this parameter
(from 10$^7$ yr to 10$^{10}$ yr) are smaller than the ones obtained
between the two metallicities considered (one can guess this small
dependence on the quenching time by noting that in Fig.\ref{figimfage}
the FUV-NUV color for a constant SFR depends little on the age).  Thus
our results for the two metallicities provide a realistic range for
the age of the event, independently of the duration of the star
formation event anterior to the quenching.
The results are
shown in Fig. \ref{figimfage}, in which the horizontal dotted lines
show the colors measured in our LSB galaxies.
Unless the IMF is severely truncated, only one galaxy is consistent
with ongoing star formation. Note that the models are dust-free,
and that the colors obtained for ongoing star formation (-0.2 to
0.2) for different metallicities and ages are consistent with
the FUV-NUV corrected for extinction of normal star forming galaxies,
as shown in the bottom panel of Fig.3.
To explain the colors of most of the LSB galaxies in the absence of
dust, we thus need to either use of a truncated IMF, or that the star
formation was quenched for a significant amount of time, between about
0.7 to 1.4 Gyr (assuming a low metallicity) or 0.1 to 0.3 Gyr (for a
solar metallicity), this time being an increasing function of the
total mass of the galaxy.
Again, even if it is quite possible that the star formation proceed by
burst episodes with quiet periods in between, or fluctuates
significantly
\citep[e.g.][]{vallenari05,boissier03,vandenhoek00,gerritsen99,oneil98},
it is somewhat strange that the time elapsed after the last burst is
longer (redder color, even if this trend is not very strong) for
galaxies with larger HI masses.

%
%
\citet{vallenari05} shows the SFR history for UGC5889, obtained from
CMD diagrams. They conclude that SFR proceeded in episodes of
very low rates ($10^{-2}$ \sfru), spaced by periods of quiescence.
However, from their Fig. 9, no period of quiescence longer than $\sim$
20 Myr occurred, concerning a dwarf LSB galaxy. It is even more difficult
to think of a reason why in more massive LSB galaxies (such as the ones
analyzed here, and in which stochasticity should have a smaller effect),
those periods could last up to $\sim$ 300 Myr.
The N-body simulations of \citet{gerritsen99} result in strongly 
fluctuating star formation histories, also with timescales of about 20 Myr.
\citet{boissier03} had also to advocate bursts and quiescent phases
to explain several observational facts in LSB galaxies, as the 
scatter in the Tully-Fisher and in the gas to luminosity ratio,
as well as the existence of some red LSB galaxies \citep{oneil97}.
According to their model, quiescent phases should be longer (up to
1 Gyr) to  explain these colors.
These works suggest that star formation in LSB galaxies or in low
density regions may proceed by bursts followed by quiescent phases,
however the timescales are not in perfect agreement with the ones we
need to explain the FUV-NUV colors.

\section{Star Formation Histories of \\ LSB galaxies}

In section \ref{secsfr}, we computed the SFR using standard calibrations.
However, the FUV-NUV color suggests either an extinction effect, a drastic
truncation of the IMF, or that the UV emitting regions are older and older
for more massive galaxies (we actually refer to the HI mass of the galaxies,
for which we have measurements).

If the effect is due to extinction, then the SFRs have been underestimated,
and the real SFRs in LSB galaxies are similar to the one in normal galaxies. However, with such 
an assumption, we would have to explain the low level of infrared emission, 
and the other signs of transparency. It would also be hard to explain the other 
signs of youth usually observed in LSB galaxies (blue colors, young ages, low metallicity).

\begin{figure*}
\includegraphics[angle=-90,width=15cm]{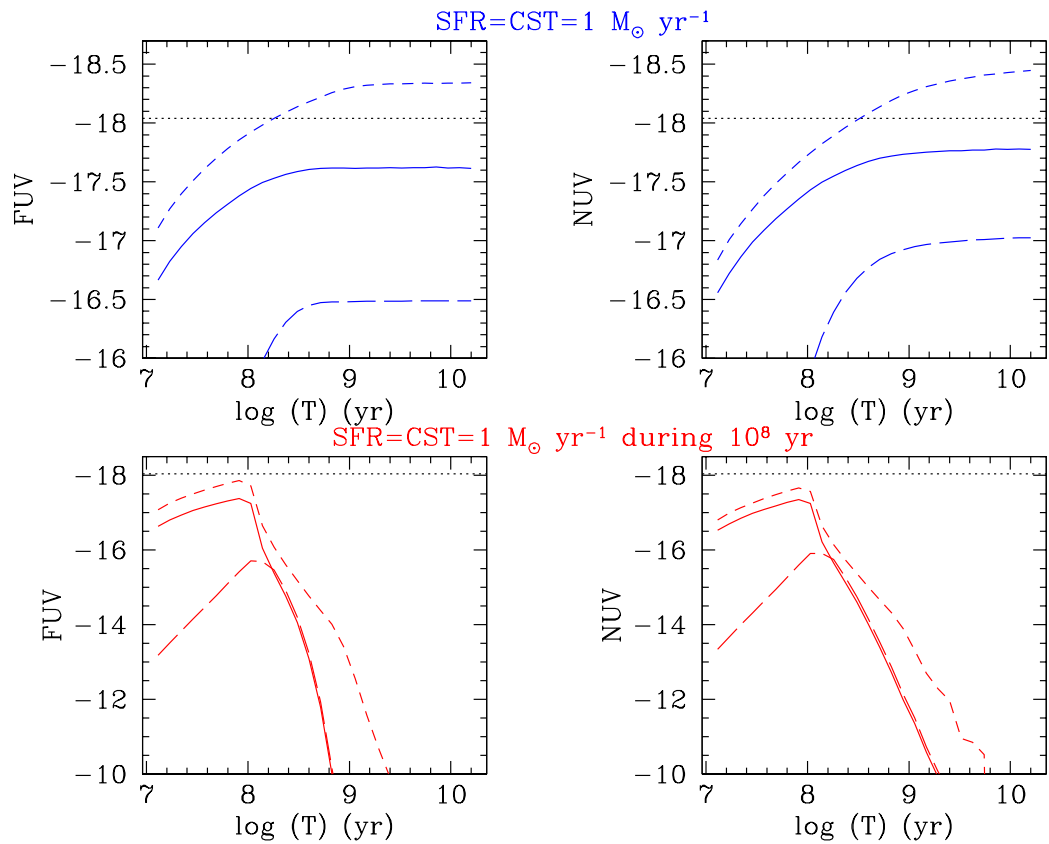}
\caption{FUV and NUV magnitudes for a star formation of 1 \sfru, continuously (top)
or quenched after 10$^8$ yr (bottom). The horizontal line shows the magnitude obtained using
the calibration of \citet{kennicutt98}. The various curves correspond to the various assumptions
concerning the IMF and the metallicity, as presented in Fig$.$~\ref{figimfage}.
\label{figsfruv}}
\end{figure*}

For the two other assumptions, we can use the same models as in section
\ref{secmodelsam}. We computed for each of them the FUV and NUV magnitudes
corresponding to star formation rates of 1 \sfru{} (continuously, or quenched
after 10$^8$ yr). The results are given in Fig. \ref{figsfruv}.

First, for a continuous star formation rate, we find that our models
are consistent with the conversion factor given by
\citet{kennicutt98}. However, depending on the metallicity, deviations
with respect to this calibration up to about $\sim$ 0.5 magnitudes are
observed (a factor $\sim$ 1.5 on the SFR).

If we adopt the severe truncation at 5 \msun{} on the IMF, we predict of course
a much weaker UV emission, by 1.5 magnitude in FUV and about 1 magnitude in 
NUV. If this is true, then our NUV SFR were underestimated by a factor about 3. 
In that case, the star formation rate at a given HI mass (or star formation
efficiency) in LSB galaxies compared to ``normal'' galaxies would be barely lower.

As mentioned above, several works
have suggested that the star formation in LSB galaxies proceeds in a
sequence of bursts and quiescent phases
\citep{vallenari05,gerritsen99,boissier03}. We try in the following to
see what we can conclude concerning the star formation history of LSB
and low density regions under this assumption.
We should also mention the study of 
\citet{boquien07} who discussed various star formation indicators
(including the UV) in low density HI probably expelled
during a galactic collision. They found that the UV actually overestimates the
current value of the SFR, but in that case the current value of the 
SFR is not very pertinent as the SFR was much higher shortly after 
the collision.

\begin{figure}
\includegraphics[width=8cm]{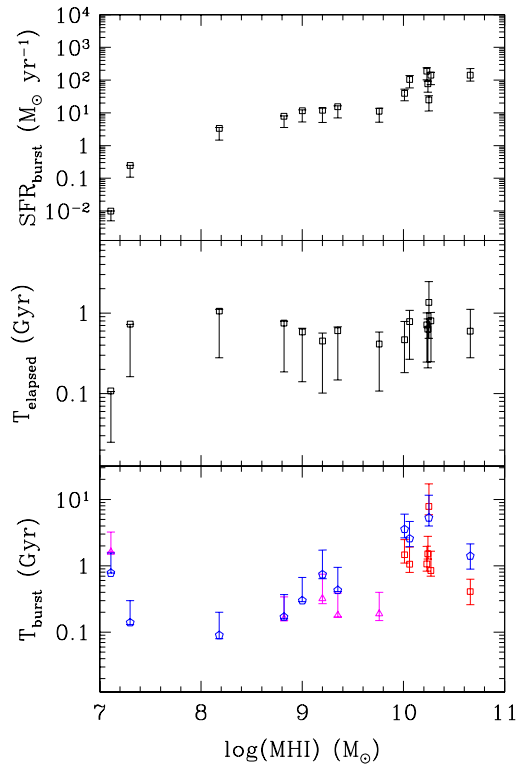}
\caption{SFR during the star formation event (top), time elapsed since
the last star formation event (middle), time necessary to build up the total
stellar mass of the galaxy (bottom) as a function of the HI mass.
The points show values derived assuming the metallicity luminosity relationship
of \citet{mcgaugh94}. The errorbars indicates values obtained when the metallicity
is moved within the range $Z_{\odot}$/20 to $Z_{\odot}$, emcompassing values found
in LSB galaxies.
In the bottom panel, squares, triangles, and pentagons correspond to adopting the
stellar masses derived from K, R and  $r$ band magnitudes respectively.
\label{figplot2go1}}
\end{figure}

Adopting the curves at the bottom of Fig. \ref{figsfruv}, and the one
in Fig. \ref{figimfage} in the case of a quenched SFR, we can compute
from the FUV-NUV color the time elapsed after the quenching, and then
from the NUV evolution given in Fig. \ref{figsfruv} the level of SFR
during the active phase. We do this for both the solar and low
metallicity (Z$_{\odot}$/20) cases in order to have an idea of the 
uncertainty due to the metallicity. 
For each galaxy, we also compute the more plausible value by
fitting the luminosity-metallicity relationship in \citet{mcgaugh94}
($log Z/Z_{\odot}=-2.81-0.11 M_B$), and use the metallicity obtained
in that way ($Z_{M_B}$). Due to the paucity of measurements of
abundances in LSB galaxies, this value is however quite uncertain.
The results are given in Table \ref{tabsfralter}.
We also show the values derived as a function of the HI mass 
(see Fig. \ref{figplot2go1}).
Our galaxies are characterized by a time elapsed since the last burst of a few
100 million years up to more than 
1 Gyr, and very high Star Formation Rates (at least for galaxies
more massive than 10$^{10}$ \msun{} in HI) during the active phase. 
The real amount of stars formed depends however
on how long this active phase lasted.


The amplitude of the bursts suggested by our 
results are quite large with respect to the one found in e.g.\citet{gerritsen99} or \citet{vallenari05} : a
few \msun{} yr$^{-1}$ for the dwarf LSB galaxies, up to several
hundred \msun{} yr$^{-1}$ in massive LSB galaxies. 
On the other hand, \citet{oneil98} show that the present SFR in LSB
galaxies is too small to produce their total luminosity, suggesting
that the SFR has been much larger in the past.
We note that some of the parameters involved in the computation of the
theoretical FUV-NUV color are uncertain. This could have some effect
on the elapsed time and burst amplitudes derived.
Also, the burst could be extended in time (instead of an instantaneous
quenching), making the evolution of the FUV luminosity shown in Fig.
\ref{figsfruv} less abrupt. This could make compatible our elapsed
time and amplitude bursts with the results of the  works of \citet{gerritsen99},
\citet{vallenari05}, or \citet{vandenhoek00} which concern relatively
low-mass LSB galaxies.  It would be hard however to diminish
sufficiently the $SFR_{burst}$ derived in giant LSB galaxies 
so to change the trends seen in Fig$.$~\ref{figsfruv}: galaxies more massive
than about 10$^{10}$ \msun (in HI) show larger SFRs and elapsed time since 
the last burst than less massive ones (even if the latter trend is unclear
when adopting the more plausible metallicity).
In this context, we should note that \citet{mapelli07} suggested that
ring-galaxies (like the Cartwheel) may evolve into giant LSB galaxies
like Malin~1 in their late stage (about 500 Myr after a collision),
what may fit with the large $SFR_{burst}$ and elapsed time that we
find.  However their analysis did not include UV data or predictions
for this wavelength.

With the ellapsed time that we find, one could expect to find almost
no H$\alpha$ emission in LSB galaxies, while it is commonly observed
\citep[e.g. references in Table \ref{tabsfr} or e.g. the rotation
curves of ][]{mcgaugh01,swaters00}.  However, many of these H$\alpha$
detections concern relatively low luminosities
\citep{swaters00} or dwarf \citep{vanzee97} LSB galaxies, while the
elapsed times we find are significantly large only in the more massive
LSB galaxies in which the presence of H$\alpha$ would pose stronger
constraints.
Also, our method based on the FUV-NUV color is
able to date the last significant event contributing to the UV
spectra.  Smaller and more recent star formation could have occurred,
as long as the UV emission is still dominated by the older event (This
would also lead to deriving lower SFR from H$\alpha$ than from the
UV).  Finally, the quenching could not be
instantaneous,  but rather present a smooth decline (we do not attempt to
model it to avoid the multiplication of free parameters),
making H$\alpha$ visible for a longer time.


In the following, we will estimate the total duration of the active
phases needed to form all the stars in these galaxies, given
$SFR_{burst}$.

Assuming an average K-band mass-to-light ratio of $log(M/L)=-0.3$
\citep{dejong01}, we can compute the stellar mass from the K band
total magnitude for galaxies with 2MASS data.  For galaxies in
\citet{deblok95}, we can compute a stellar mass from the R band
magnitude, with the stellar mass-to-light ratio of \citet{dejong01}
depending on B-R.
From the SDSS $r$-band data, we can also compute a stellar mass
adopting the average trend shown in \cite{kauffmann03}.  These stellar
masses are given in Table \ref{tabstelmass}.  They are rough estimates,
with the differences between the various masses giving an idea of the
uncertainty ($\sim$ 0.2 dex).

Assuming that the star formation proceed by bursts of amplitude
$SFR_{burst}$ (Table \ref{tabsfralter}), we determined 
how long it takes to produce the whole stellar mass of the galaxy
($T_{burst}=M_*/(1-R) SFR_{burst}$), where R is the returned fraction,
for which we adopt R=0.3, corresponding to the \citet{kroupa93} IMF.
The values obtained (for each estimation of the stellar mass, and for
the SFR determined for various metallicities) are given in Table
\ref{tabstelmass} and Fig. \ref{figplot2go1}. The ``burst'' time is
actually the total time during which the galaxy had to be in an active
phase to build up its stellar mass (assuming no stars at all were formed
during the quiescent periods), even if it may have been split in
several events of similar $SFR_{burst}$.

From Fig.  \ref{figplot2go1}, we can see that the results are quite
dispersed, due to the approximations we have to make to derive SFR and
stellar masses. However, it seems that above 10$^{10}$ solar masses
(corresponding to galaxies in which we find very high $SFR_{burst}$),
the time elapsed after the last burst is increasing with the mass. At
the same time, $T_{burst}$, the time during which the galaxy was in a
``burst'' phase during its history, decreases with the mass (this
result is however very uncertain, especially due to the uncertainty on
the metallicity, affecting $T_{elapsed}$ and the quantities derived
from it).
This opposite behavior of $T_{elapsed}$ and $T_{burst}$ is indeed
expected when the fraction of ``active phase'' duration with respect
to the life of galaxy gets smaller, there is less and less chance to
catch the galaxy during this phase (for galaxies with $T_{burst} \sim$
1 Gyr, chances are already lower than 10 \%); and more and more
chances to find a galaxy with larger elapsed time.

\section{Conclusions}

In this paper, we presented the GALEX UV observations of 18 low surface 
brightness galaxies, with known HI content. The UV light relatively to
the optical one is more extended than in normal star forming galaxies.
Morphologies similar to extended UV disks (XUV) are often found 
(about 70 \% of our objects, F568-3 (Malin~2) is
a very nice example), although not systematically. 

Adopting standard calibrations to convert the UV light into star
formation rates, we obtain a large range of SFR (a few 10$^{-3}$ to a
few solar masses per year), depending on the HI mass of the galaxy.
Massive LSB galaxies have lower SFR than normal star forming galaxies
with the same gas reservoir (by a factor $\sim$ 5). Such a lower
efficiency for forming stars is expected in various models, where e.g.
LSB galaxies have larger radial extent due to larger spin parameters.
However, the SFR obtained in such a way are highly uncertain, due to
the very red FUV-NUV color (especially in massive LSB galaxies)
that we measure, that might indicate a
highly non-constant star formation history.

Several interpretations for this color are possible, and
more work on stellar populations in LSB galaxies is needed to obtain a
definitive answer.  The various possibilities we considered are :

i) A dust attenuation effect. 
Note, however, that the low infrared emission found in LSB galaxies
makes this explanation unlikely.

ii) Variations in the Initial Mass Function.  The FUV-NUV colors can
be recovered in a constant star-formation scenario, but only for
extremely steep (or truncated) IMF. Moreover, we find that the more
massive galaxies (as measured with the HI mass) are redder, while the
IMF proposed \citet{weidner05}, depending on the star formation rate,
should have the opposite effect. 
We do not exclude the possibility that a IMF effect might play a role
in galaxies with low HI masses (and thus low star formation rates), but
it cannot be responsible for the trends seen in the more massive
galaxies.

iii) An age-effect. Red FUV-NUV colors can be obtained if the SFR was
quenched. In this scenario, we find that the more massive LSB galaxies
have known more massive star formation event (larger $SFR_{burst}$,
followed by relatively long quiescent phases of several hundred million
years). If these events are spread over the whole history of the
galaxy, on average the SFR can be about constant over the Gyr
time-scale, giving as a result optical-near infrared colors similar to
late type galaxies (as those observed), while the FUV-NUV can get
relatively red during the quiescent phase, dominating the lifetime of
the galaxy. To explain the fact that most massive galaxies are redder,
it would be necessary to assume that stars in these galaxies are
formed during bursts of larger amplitudes, separated by longer
quiescent epochs. The more extreme example of this effect is UGC~00568
for which the quiescent phase could have lasted up to 2 Gyr (also
affecting the optical colors in this case: indeed its SED is typical
of early type galaxies), while a huge amount of HI is present.
A clear physical reason for this is still to be found.
We can only speculate that massive LSB galaxies have large gas
reservoirs, in which star formation is suddenly turned on 
only occasionally.
The next large burst event cannot occur until the gas
reservoir has been built again. 

In any case, it seems quite dangerous to derive SFR from the UV light
in low density regions (LSB galaxies, outer XUV disks) without taking
some precautions, especially if red FUV-NUV colors are observed or if
there is no independent way to date the star formation event.

\acknowledgments

We thank the referee, Stacy McGaugh, for his very constructive comments,
as well as David Thilker and Ted Wyder for discussions.

This research is first based on the GALEX GI program GALEXGII04-0000-0067,
as well as public archival data, available at the  MAST archive.

This research has made use of the NASA/IPAC Extragalactic Database
(NED) which is operated by the Jet Propulsion Laboratory, California
Institute of Technology, under contract with the National Aeronautics
and Space Administration.

We acknowledge the usage of the HyperLeda database (http://leda.univ-lyon1.fr).

This research has made use of the GOLD Mine Database. 

{\it Facilities:} \facility{GALEX}.



\clearpage
\begin{deluxetable}{l r r r r r r } 
\tablecaption{Time elapsed since last burst and Star Formation Rates during the burst  \label{tabsfralter}} 
\startdata 
        Name     & \multicolumn{3}{c}{------ Time Ellapsed ------} & \multicolumn{3}{c}{--------- SFR $_{burst}$ ---------}  \\ 
                 & \multicolumn{3}{c}{(Gyr)} & \multicolumn{3}{c}{(\sfru)}  \\ 
                 & Low Z  &  $Z_{M_B}$ &  high $Z$  & Low $Z$  &  $Z_{M_B}$ &  high Z   \\  
        UGC00568 &   2.453 &   1.353 &   0.487 &  33.327 &  24.947 &  11.415 \\ 
        UGC01230 &   0.583 &   0.413 &   0.108 &  13.873 &  11.166 &   5.196 \\ 
        UGC02936 & --- & --- & --- & ---  & --- & --- \\ 
     OBC97-C04-2 &   1.114 &   1.057 &   0.278 &   3.550 &   3.362 &   1.479 \\ 
         F564-V3 &   0.106 &   0.108 &   0.025 &   0.010 &   0.010 &   0.005 \\ 
        UGC05209 &   0.724 &   0.732 &   0.163 &   0.251 &   0.245 &   0.109 \\ 
          F568-1 &   0.674 &   0.607 &   0.148 &  16.437 &  15.530 &   7.032 \\ 
          F568-3 &   0.565 &   0.449 &   0.102 &  13.921 &  11.947 &   5.125 \\ 
        UGC05750 &   0.650 &   0.583 &   0.141 &  12.367 &  11.587 &   5.262 \\ 
       PGC135754 &   1.080 &   0.782 &   0.267 & 139.026 & 105.025 &  57.804 \\ 
          F568-6 & --- & --- & --- & ---  & --- & --- \\ 
         F571-V1 &   0.794 &   0.752 &   0.186 &   8.056 &   7.818 &   3.570 \\ 
          Malin1 &   1.114 &   0.597 &   0.278 & 225.283 & 142.340 &  93.836 \\ 
        PGC45080 & --- & --- & --- & ---  & --- & --- \\ 
          F530-1 &   1.016 &   0.799 &   0.249 & 174.621 & 143.097 &  73.402 \\ 
          F533-3 &   0.847 &   0.630 &   0.209 &  93.536 &  77.953 &  42.575 \\ 
         NGC7589 &   0.785 &   0.466 &   0.182 &  52.945 &  39.502 &  23.342 \\ 
        PGC71626 &   1.005 &   0.716 &   0.247 & 241.341 & 187.828 & 102.281 \\ 
\enddata 
\tablecomments{$Z$ indicates the metallicity adopted. High and Low metallicity are respectively Solar and 1/20 Solar metallicities. $Z_{M_B}$ is the metallicity deduced from the Metallicity-Luminosity relationship in \citet{mcgaugh94}.}
\end{deluxetable} 

\clearpage

\begin{deluxetable}{l r r r r r r r r r r r r r r  } 
\rotate 
\tablecaption{Stellar mass and active phase duration  \label{tabstelmass}} 
\startdata 
              &   \multicolumn{4}{c}{------------K band------------} &    \multicolumn{4}{c}{------------R band------------} &  \multicolumn{4}{c}{------------r band------------}  \\
        Name     & {\small log(M$_*$(K))} & \multicolumn{3}{c}{----- $T_{burst}$ -----} & {\small log(M$_*$(R))}  & \multicolumn{3}{c}{----- $T_{burst}$ ----- }   & {\small log(M$_*$(r))}  & \multicolumn{3}{c}{----- $T_{burst}$ ----- } \\
                 &   \msun & \multicolumn{3}{c}{Gyr} &   \msun &  \multicolumn{3}{c}{Gyr} &   \msun &  \multicolumn{3}{c}{Gyr} \\  
            &   & $Z_{\odot}/20 $ & $Z_{M_B}$ & $Z_{\odot}$ &   & $Z_{\odot}/20 $ & $Z_{M_B}$ & $Z_{\odot}$  &   & $Z_{\odot}/20 $ & $Z_{M_B}$ & $Z_{\odot}$ \\ 
        UGC00568  &   11.14 &    5.88 &    7.86 &   17.18 &  ---& --- & ---   &  ---   &  10.97 &    3.97 &    5.30 &   11.58 \\ 
        UGC01230  &  --- & --- & ---  &  ---  &   9.16 &    0.15 &    0.19 &    0.40 &  --- & --- & --- & --- \\ 
        UGC02936  &  --- & --- & ---  &  ---  & ---& --- & ---   &  ---   & --- & --- & --- & --- \\ 
     OBC97-C04-2  &  --- & --- & ---  &  ---  & ---& --- & ---   &  ---   &   8.32 &    0.08 &    0.09 &    0.20 \\ 
         F564-V3  &  --- & --- & ---  &  ---  &   7.05 &    1.61 &    1.61 &    3.23 &    6.74 &    0.78 &    0.78 &    1.56 \\ 
        UGC05209  &  --- & --- & ---  &  ---  & ---& --- & ---   &  ---   &   7.37 &    0.13 &    0.14 &    0.30 \\ 
          F568-1  &  --- & --- & ---  &  ---  &   9.28 &    0.17 &    0.18 &    0.39 &    9.67 &    0.41 &    0.43 &    0.95 \\ 
          F568-3  &  --- & --- & ---  &  ---  &   9.43 &    0.27 &    0.32 &    0.75 &    9.79 &    0.64 &    0.74 &    1.73 \\ 
        UGC05750  &  --- & --- & ---  &  ---  & ---& --- & ---   &  ---   &   9.39 &    0.29 &    0.30 &    0.67 \\ 
       PGC135754  &   10.89 &    0.80 &    1.06 &    1.92 &  ---& --- & ---   &  ---   &  11.28 &    1.94 &    2.57 &    4.66 \\ 
          F568-6  &  --- & --- & ---  &  ---  & ---& --- & ---   &  ---   & --- & --- & --- & --- \\ 
         F571-V1  &  --- & --- & ---  &  ---  &   8.93 &    0.15 &    0.16 &    0.34 &    8.96 &    0.16 &    0.17 &    0.37 \\ 
          Malin1  &   10.62 &    0.26 &    0.41 &    0.63 &  ---& --- & ---   &  ---   &  11.15 &    0.89 &    1.41 &    2.14 \\ 
        PGC45080  &  --- & --- & ---  &  ---  & ---& --- & ---   &  ---   & --- & --- & --- & --- \\ 
          F530-1  &   10.93 &    0.70 &    0.85 &    1.66 &  ---& --- & ---   &  ---   & --- & --- & --- & --- \\ 
          F533-3  &   10.92 &    1.27 &    1.52 &    2.78 &  ---& --- & ---   &  ---   & --- & --- & --- & --- \\ 
         NGC7589  &   10.61 &    1.10 &    1.47 &    2.49 &  ---& --- & ---   &  ---   &  10.99 &    2.65 &    3.55 &    6.01 \\ 
        PGC71626  &   11.15 &    0.83 &    1.07 &    1.97 &  ---& --- & ---   &  ---   & --- & --- & --- & --- \\ 
\enddata 
\tablecomments{Stellar mass derived from various photometric bands (see text), and burst duration obtained from it, adopting various metallicities.}
\end{deluxetable}

\clearpage

\appendix

\section{Individual figures (online only)}

\begin{figure}
\includegraphics[angle=90,width=19cm]{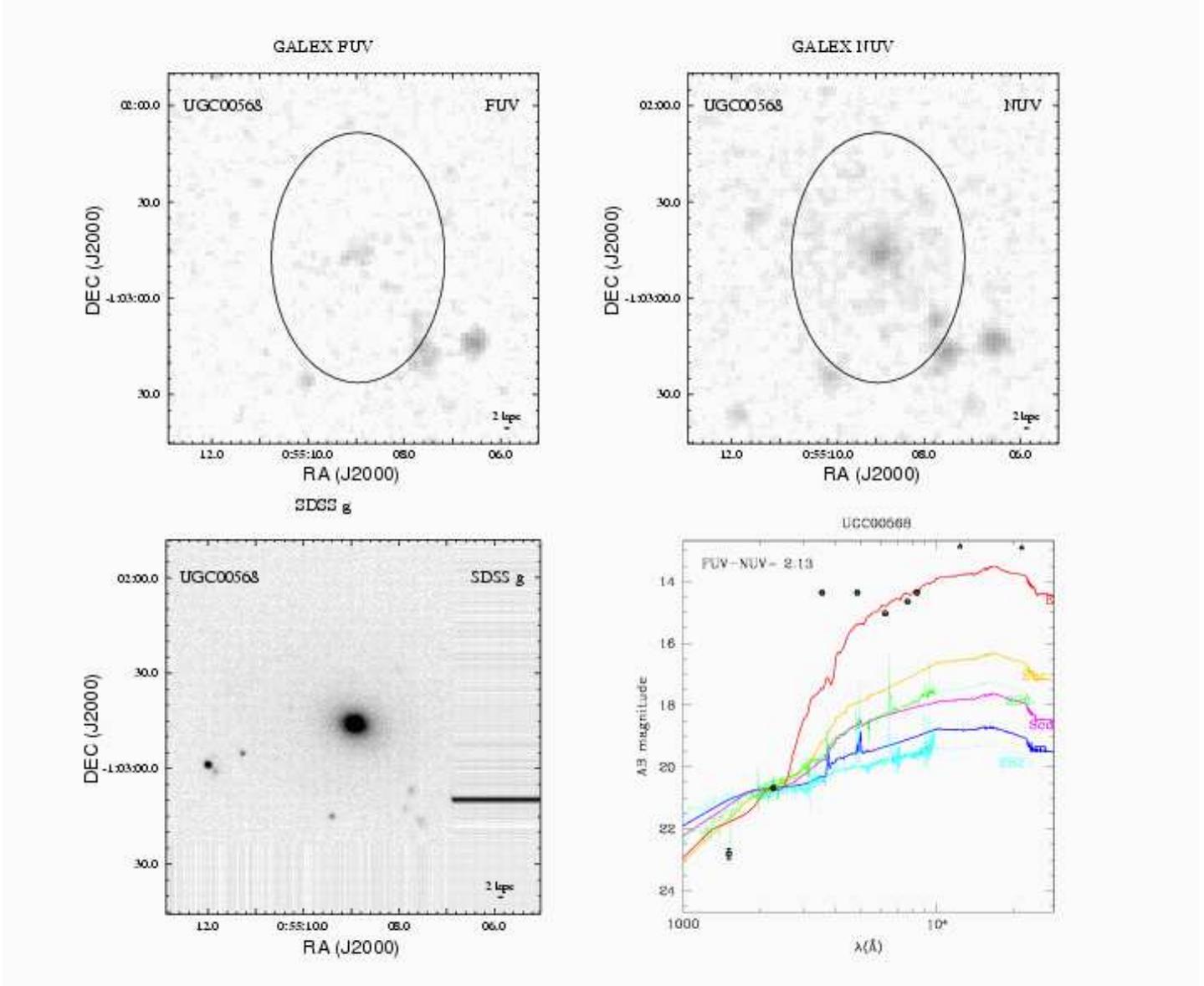}
\figurenum{9}

\caption{ THIS PREPRINT ONLY INCLUDES ONE EXEMPLE FOR FIGURE SET  9.
  COMPLETE FIGURE SET WILL BE AVAILABLE ELECTRONICALLY ON THE JOURNAL
  SITE AND AT :
  \emph{http://www.oamp.fr/people/boissier/preprint/}. Images
  and SED (bottom-right). GALEX and SDSS magnitudes were computed in
  this paper, other wavelengths are taken from NED and the literature
  (see text). The name of the galaxy is indicated in each panel.
  Ellipses indicate the optical diameter (2a), for the adopted
  inclination and PA (see parameters in table 1). In the bottom-right
  panel, circles with error-bars show our photometry of GALEX and SDSS
  data. Triangles correspond to data from 2MASS, squares to values
  from \citet{hunter06}, crosses to values from \citet{deblok95},
  diamonds to values from \citet{mcgaugh94b}.  Templates in the
  bottom-right panel are taken from Coleman et al.  (1980; with the
  extrapolation of Arnouts et al. 1999) and \citet{kinney93}, as
  distributed in Le Phare
  (http://www.lam.oamp.fr/arnouts/LE\_PHARE.html)
  \label{figindividual}}
\end{figure}

Fig. \ref{figindividual} shows the GALEX FUV and NUV images, as well
as the SDSS $g$ image when available (replaced by the DSS-1 red image when
this is not the case). The bottom-right panel shows the broad band SED
(in AB magnitudes) constructed from 
UV-optical-near-infrared photometry when available, including  
GALEX UV, optical data from the SDSS and the literature, and 2MASS near-infrared
(see text).

\clearpage

\end{document}